\documentclass[10pt]{revtex4}
\usepackage{amssymb}
\usepackage{latexsym}
\usepackage{epsfig}
\begin{document}
 
\title{Quantum spectrum of tachyonic black holes in a brane-anti-brane system }

\author{ Aroonkumar Beesham$^{1}$\footnote{beesham@mut.ac.za}}
\affiliation{$^{1}$Faculty of Natural Sciences, Mangosuthu University of Technology, P O Box 12363, Umlazi 4026,  South Africa }

\begin{abstract}

\textbf{Abstract:} 

Recently, some authors have considered  the quantum spectrum of black holes . This consideration is extended to tachyonic black holes in a brane-anti-brane system.  In this study, black holes are constructed from two branes which are connected by a tachyonic tube. As the branes come closer to each other, they evolve and  make a transition to thermal black branes.   It will be shown that the spectrum of these black holes depends on the tachonic potential and the separation distance between the branes. By decreasing the separation distance, more energy  emerges and the spectrum of the black hole increases.
\vspace{5mm}\noindent\vspace{0.8mm}\newline \textbf{Keywords:} : Black hole, quantum spectrum, Shape, Tachyon, Branes
\end{abstract}

\maketitle

\section{Introduction:} 

Of late, some scientists suggested a more exact black hole
effective temperature with reference to the quantum spectrum of  black holes     \cite{w1,w2}. This  temperature includes   both the non-thermal Hawking
radiation and
 the radiation of subsequent Hawking quanta. In \cite{w1,w2}, it was shown that the  quantization depends on  the quantum  quasi-normal modes of the black hole, but there were certain approximations implicitly made in those calculations. In \cite{w3}, Corda extended the previous calculations by removing these approximations, and obtained  corrected expressions for the quantization 
and thereby also for the Bekenstein-Hawking entropy. Other researchers \cite{ww1,ww2} using different methods have also obtained  the black hole spectrum.  Motivated  by these works, we consider the quantum spectrum of tachyonic black holes in a brane-anti-brane system. These black holes are constructed from a pair of branes and anti-branes which are connected by a tachyonic tube \cite{w4,w5,w6}. By decreasing the separation distance between branes, the tachyonic potential between them grows and the tachyonic black holes emit more spectra.

The outline of this paper is as follows: In section 2, we calculate the quantum spectrum for  tachyonic black holes which are constructed from a brane, an anti-brane and a tachyonic tube.   In section 3, we generalize this discussion to thermal black holes.  The last section is devoted to a summary and conclusion.

\section{The quantum spectrum for tachyonic black holes:} 

In this section we will firstly consider a system of a brane and an anti-brane which are connected by a tachyonic tube. By increasing the tachyonic potential between the branes, this system evolves to a black hole. We will calculate the spectrum of this tachyonic black hole. In \cite{w3}, it was shown that the entropy for the  black hole is given by
\begin{eqnarray}
&& S_{BH}=4\pi \Big(M^{2}-\frac{n}{2}\Big)\label{E}
\end{eqnarray}
where n is the number of quantum states and M is the energy of the black hole. Now, we wish to calculate the energy of the tachyonic black hole.  For a black hole in the brane-anti-brane system, the total potential energy can be obtained by summing over the potentials of the branes and the spaces between them:
\begin{eqnarray}
 V_{tot}=V_{brane} + V_{extra}\label{po1}
\end{eqnarray}
The extra potential is a function of the fields  which can move between the branes. Such fields transmit forces between the branes, playing a major role in the evolution of the black holes located on the branes. These fields turn out to be tachyons.

\begin{figure*}[thbp]
	\begin{center}
		\begin{tabular}{rl}
			\includegraphics[width=8cm]{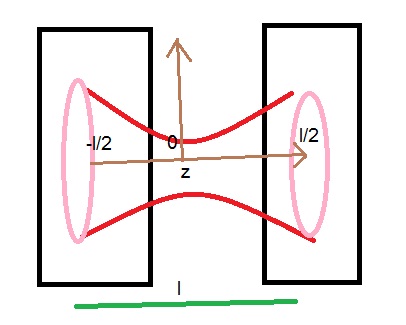}
		\end{tabular}
	\end{center}
	\caption{Pair of
$D3$-$\overline{D3}$-brane pairs 
at $z_{1} = l/2$ and $z_{2} = -l/2$}
\end{figure*}

 To build a tachyonic black hole in this theory and calculate the tachyonic potential,  consider a set of
$D3$-$\overline{D3}$-brane pairs  situated at  $z_{1} = l/2$ and $z_{2} = -l/2$ respectively as shown in figure 1. $z$ is the transverse coordinate to the branes and $\sigma$ is the radius on the world-volume.  The induced metric on the
brane is:
\begin{eqnarray}
\gamma_{ab}d\sigma^{a}d\sigma^{b} = -d\tau^{2} + (1 +
z'(\sigma)^{2})d\sigma^{2} + \sigma^{2}(d\theta^{2} +
\sin^{2}\theta d\phi^{2}) \label{Q5}
\end{eqnarray}

 For
the  case of a single $D3$-$\overline{D3}$-brane pair with
open string tachyon, the action is \cite{q6,q5}:
 \begin{eqnarray}
&& S_{tot-extra}=-\tau_{3}\int d^{9}\sigma \sum_{i=1}^{2}
V(TA,l)e^{-\phi}(\sqrt{-det A_{i}})\nonumber \\&&
(A_{i})_{ab}=\left(g_{MN}-\frac{TA^{2}l^{2}}{Q}g_{Mz}g_{zN}\right)\partial_{a}x^{M}_{i}\partial_{b}x^{M}_{i}
+F^{i}_{ab}+\frac{1}{2Q}((D_{a}TA)(D_{b}TA)^{\ast}+(D_{a}TA)^{\ast}(D_{b}TA))\nonumber
\\&&
+il(g_{az}+\partial_{a}z_{i}g_{zz})(TA(D_{b}TA)^{\ast}-TA^{\ast}(D_{b}TA))+
il(TA(D_{a}TA)^{\ast}-TA^{\ast}(D_{a}TA))(g_{bz}+\partial_{b}z_{i}g_{zz}),
\label{Q27}
\end{eqnarray}
where
  \begin{eqnarray}
&& Q=1+TA^{2}l^{2}g_{zz}, \nonumber \\&&
D_{a}TA=\partial_{a}TA-i(A_{2,a}-A_{1,a})TA,
V(TA,l)=g_{s}V(TA)\sqrt{Q}, \nonumber \\&& e^{\phi}=g_{s}( 1 +
\frac{R^{4}}{z^{4}} )^{-\frac{1}{2}}, \label{Q28}
\end{eqnarray}
The quantities $\phi$, $A_{2,a}$ and $F^{i}_{ab}$ are the dilaton field, gauge
fields and field strength, respectively, on the world-volume of the non-BPS
brane.  $TA$ is the tachyon field, $\tau_{3}$  the brane tension
and $V(TA)$  the tachyon potential. Indices $a,b$ stand for the
tangent directions of the $D$-branes, whereas the indices $M,N$ run over the
background ten-dimensional space-time directions. The indices  $i$ = 1 and 2 represent the $Dp$-brane and
the anti-$Dp$-brane, respectively. Then
the distance between the $D$-branes is given by $z_{2} - z_{1}
= l$. In the above action, we use units such that 
$2\pi\acute{\alpha}=1$.

In writing the action of the D3-brane, we assume that  that  $\sigma$ is dependant only on the tachyon field $TA$,  and that the
gauge fields are zero.  Thus  in the
region  $r> R$ and $TA'\sim constant$, the action (\ref{Q27}) is
  \begin{eqnarray}
S_{D3} \simeq-\frac{\tau_{3}}{g_{s}}\int dt \int d\sigma \sigma^{2}
V(TA)(\sqrt{D_{1,TA}}+\sqrt{D_{2,TA}}), \label{Q29}
\end{eqnarray}
where $D_{1,TA} = D_{2,TA}\equiv D_{TA}$,
${\displaystyle V_{3}=\frac{4\pi^{2}}{3}}$ is the volume of a unit sphere $S^{3}$ and
 \begin{eqnarray}
D_{TA} = 1 + \frac{l'(\sigma)^{2}}{4}+ TA^{2}l^{2}, \label{Q30}
\end{eqnarray}
where a prime  denotes a derivative with respect to
$\sigma$. We make use of the potential
\cite{q7,q8,q9}:
 \begin{eqnarray}
V(TA)=\frac{\tau_{3}}{\cosh\sqrt{\pi}TA}. \label{Q31}
\end{eqnarray}
To calculate the energy momentum tensor, we have to take the functional derivative of the action with respect to 
metric $g_{MN}$, i.e., ${\displaystyle T^{MN} =
\frac{2}{\sqrt{-det g}}\frac{\delta S}{\delta g_{MN}}}$. We get
\cite{w5,w6},
 \begin{eqnarray}
&& T^{00}_{i,brane}=V(TA)\sqrt{D_{TA}},  \label{Q32}
\end{eqnarray}

After doing some calculations and using some approximations, we obtain:

\begin{eqnarray}
&& T^{00}_{i,brane}= \tau_{3} + V_{brane},  \label{moment1}
\end{eqnarray}
where 
\begin{eqnarray}
&&   V_{brane}= \tau_{3}[\frac{\sqrt{\pi}TA}{2}][1+ e^{-2\sqrt{\pi}TA}]^{-1}\times \nonumber
\\&& [\frac{l'(\sigma)^{2}}{4}+ TA^{2}l^{2}] \label{pot2}
\end{eqnarray}
This potential depends on the distance between the two branes and on the tachyon. the effects of the other branes has to be taken into account to get the change of the parameters with time.  It will  shown that as the branes approach each other, the tachons generate a wormhole connecting the branes which then transmits energy into the black hole from the extra dimensions. 

Thus far, we have assumed that the tachyon field changes slowly ($TA \sim t^{4}/t^{3}= t$), whilst neglacting 
${\displaystyle TA'=\frac{\partial TA}{\partial \sigma}}$ and
${\displaystyle \dot{TA}=\frac{\partial TA}{\partial t}}$. Now, we consider the tachyon field to be changing rapidly as the distance  between the brane and antibrane decreases.  So we cannot neglect $TA'$ and $\dot{TA}$. A  new wormhole forms. 
 During this time, the black hole changes from a non-phantom phase to a new
phantom phase. Thus, the phantom-dominated era of the
black hole accelerates, ending up in a  big-rip singularity. In such a
case, the action (\ref{Q27}) becomes:
  \begin{eqnarray}
L \simeq-\frac{\tau_{3}}{g_{s}} \int d\sigma \sigma^{2}
V(TA)(\sqrt{D_{1,TA}}+\sqrt{D_{2,TA}}), \label{Q35}
\end{eqnarray}
where
 \begin{eqnarray}
D_{1,TA} = D_{2,TA}\equiv D_{TA} = 1 + \frac{l'(\sigma)^{2}}{4}+ 
\dot{TA}^{2} -  TA'^{2}+ TA^{2}l^{2}, \label{Q36}
\end{eqnarray}
and where it is  assumed that $TA l\ll TA'$. Next, the Hamiltonian
related to the above Lagrangian is studied.
 The canonical momentum density is needed to derive the Hamiltonian, i.e.,
 ${\displaystyle \Pi =
\frac{\partial L}{\partial \dot{TA}}}$ associated with the tachyon, that is
 \begin{eqnarray}
\Pi = \frac{V(TA)\dot{TA}}{ \sqrt{1 + \frac{l'(\sigma)^{2}}{4}+
\dot{TA}^{2} -  TA'^{2}}}, \label{Q37}
\end{eqnarray}
and the Hamiltonian is:
\begin{eqnarray}
H_{DBI} = 4\pi\int d\sigma  \sigma^{2} \Pi \dot{TA} - L.
 \label{Q38}
\end{eqnarray}

Now we choose $\dot{TA} = 2 TA'$, and obtain:
\begin{eqnarray}
H_{DBI} = 4\pi\int d\sigma \sigma^{2} \left[\Pi
(\dot{TA}-\frac{1}{2}TA')\right] + \frac{1}{2}TA\partial_{\sigma}(\Pi
\sigma^{2}) - L
 \label{Q39}
\end{eqnarray}
In the second step of the above  equation, we have integrated the term proportional to $\dot{TA}$ by parts. This  indicates that  the tachyon can
be studied as a Lagrange multiplier by imposing the constraint
$\partial_{\sigma}(\Pi \sigma^{2}V(TA))=0$ on the canonical
momentum. By solving the above equation, we get:
\begin{eqnarray}
\Pi =\frac{\beta}{4\pi \sigma^{2}},
 \label{Q40}
\end{eqnarray}
where $\beta =$ constant. By (\ref{Q38}) and (\ref{Q40}),  we
get:
\begin{eqnarray}
&& H_{DBI} = \int d\sigma V(TA)\sqrt{1 + \frac{l'(\sigma)^{2}}{4}
+ \dot{TA}^{2}+ TA^{2}l^{2}}F_{DBI} ,  \nonumber \\&&
F_{DBI}=\sigma^{2}\sqrt{1 + \frac{\beta}{\sigma^{4}}}\label{Q41}
\end{eqnarray}
We then vary
(\ref{Q41}), and calculate the equation of motion  for $l(\sigma)$:
\begin{eqnarray}
&&\left(\frac{l'F_{DBI}}{4\sqrt{1+
\frac{l'(\sigma)^{2}}{4}}}\right)'=0\label{Q42}
\end{eqnarray}
The solution to this equation is:
\begin{eqnarray}
&&l(\sigma) = 4\int_{\sigma}^{\infty} d\sigma
\left(\frac{F_{DBI}(\sigma)}{F_{DBI}(\sigma_{0})}-1\right)^{-\frac{1}{2}}=4\int_{\sigma}^{\infty}
d\sigma'\left(\frac{\sqrt{\sigma_{0}^{4}+\beta^{2}}}{\sqrt{\sigma'^{4}-\sigma_{0}^{4}}}\right)
\label{Q43}
\end{eqnarray}
This solution represents  a wormhole with a finite size throat for non-zero $\sigma_{0}$,   
 (See figure 1). Using equations (\ref{Q37},\ref{Q40}and \ref{Q43}) and assuming that $\dot{TA}^{2} = TA'^{2}$, we obtain:

\begin{eqnarray}
&& TA\sim\int d\sigma [\frac{\beta}{4\pi \sigma^{2}}][\sqrt{1 + [\left(\frac{\sqrt{\sigma_{0}^{4}+\beta^{2}}}{\sqrt{\sigma'^{4}-\sigma_{0}^{4}}}\right)]^{2}}]
\label{tachyon2}
\end{eqnarray}

We see from this that the tachyons depend on the coordinates of the branes and  the size of the throat of the wormhole.  By decreasing the distance  between the branes, the tachyons expand and more energy is transmitted from the extra dimensions into the brane and thus the black hole expands.  

The potential between branes can be obtained from equation (\ref{Q41}): 
\begin{eqnarray}
&& H_{DBI} = T + V_{tot} \nonumber \\&& V_{tot}\simeq \frac{3\tau_{3}}{\sigma^{3}}\int d\sigma V_{brane} F_{DBI} ,  \nonumber \\&&
F_{DBI}=\sigma^{2}\sqrt{1 + \frac{\beta}{\sigma^{4}}}\label{po3}
\end{eqnarray}
The energy density may be calculated from equations (\ref{Q41} and \ref{po3} ):

\begin{eqnarray}
&& T^{00}_{i,brane}= \int d \sigma T^{00}_{i,brane + extra}=  \int d \sigma V(TA)\sqrt{D_{TA}}F_{DBI} \simeq  V_{tot} \label{moment2}
\end{eqnarray}
where $ T^{00}_{i,brane}$ is the energy of the brane and $T^{00}_{i,brane + extra}$ is the energy of the brane-anti-brane and tube. Putting the energy density in equation (\ref{moment2}) equal to the energy density in equation (\ref{moment1}), we obtain:

\begin{eqnarray}
&&  \tau_{3} + V_{brane} =  V_{tot} ,\nonumber \\&& \tau_{3} + V_{brane} = \frac{3\tau_{3}}{\sigma^{3}}\int d\sigma V_{brane} \sigma^{2}\sqrt{1 + \frac{\beta}{\sigma^{4}}} \label{equal}
\end{eqnarray}

By multiplying equation (\ref{equal}) by
$\sigma^{3}$, 
and by differentiating  with respect to the cosmic time, we get:

\begin{eqnarray}
&& 3\sigma^{2}\dot{\sigma}V_{brane}  +\sigma^{3}\dot{V}=-3\tau_{3}\sigma^{2}\dot{\sigma} +3\tau_{3}\sigma^{2}V_{brane}\dot{\sigma}\sqrt{1 + \frac{\beta}{\sigma^{4}}}  \label{equation}
\end{eqnarray}

For $\beta\ll1$, we obtain:

\begin{eqnarray}
&&  \frac{\dot{\sigma}}{\sigma}=\frac{-\dot{V}_{brane}}{3\tau_{3} +[3 - 3\tau_{3}]V_{brane}}\label{Hubble}
\end{eqnarray}
 
Solving equations (\ref{pot2}, \ref{Q43},  \ref{tachyon2},  \ref{equal} and \ref{Hubble}) simultaneously, we obtain:

\begin{eqnarray}
 &&   V_{tot}= \tau_{3} +\tau_{3}[\frac{\sqrt{\pi}[\frac{\beta}{4\pi [t_{0}^{2}-t^{2}}][\sqrt{1 + [\left(\frac{\sqrt{t_{0}^{4}+\beta^{2}}}{\sqrt{t^{4}-t_{0}^{4}}}\right)]^{2}}]}{2}]\times \nonumber
\\&&[1+ e^{-2\sqrt{\pi}[\frac{\beta}{4\pi [t_{0}^{2}-t^{2}}][\sqrt{1 + [\left(\frac{\sqrt{t_{0}^{4}+\beta^{2}}}{\sqrt{t^{4}-t_{0}^{4}}}\right)]^{2}}]}]^{-1}\times \nonumber
\\&& [\frac{1}{4}[\left(\frac{\sqrt{t_{0}^{4}+\beta^{2}}}{\sqrt{t_{0}^{4}-t^{4}}}\right)
]+ \nonumber [\frac{\beta}{4\pi [t_{0}^{2}-t^{2}}][\sqrt{1 + [\left(\frac{\sqrt{t_{0}^{4}+\beta^{2}}}{\sqrt{t^{4}-t_{0}^{4}}}\right)]^{2}}]^{2} \times\nonumber\\&&[\left(\frac{\sqrt{t_{0}^{4}+\beta^{2}}}{\sqrt{t_{0}^{4}-t^{4}}}\right)
]^{2}] \label{pot2}
\end{eqnarray}

Substituting energy (\ref{pot2}) in equation (\ref{E}), we obtain:

\begin{eqnarray}
 && S_{BH}=4\pi \Big((V_{tot})^{2}-\frac{n}{2}\Big)
\nonumber\\&& T_{tot}=\frac{1}{4\pi V_{tot}}
\label{aEQ13}
\end{eqnarray}

The above equations show that by decreasing the distance between branes, the tachyonic energy increases. This causes  the quantum spectrum of the black hole to grow and the entropy increases.

\section{The  quantum spectrum of thermal tachyonic black branes :} 

In this section, we will generalize the method in the previous section to thermal black branes. We will show that branes move with high acceleration towards each other. This acceleration produces a curved space-time and a creates a horizon around the  system. This causes the system to evolve and make a transition to a system of black branes.

To achieve  these aims, we begin with the equation of motion for the tachyons as follows:

\begin{eqnarray}
&& -\frac{\partial^{2} TA}{\partial \tau^{2}} + \frac{\partial^{2}
TA}{\partial \sigma^{2}}=0 \label{qs1}
\end{eqnarray}
By using (\ref{Hubble}), we can write the  following re-parameterizations
\begin{eqnarray}
&& \rho = \rho_{0} \dot{\sigma}^{2}= \rho_{0}\frac{\sigma^{2}}{w^{2}} ,  \nonumber \\&& w=
\frac{3\tau_{3} +[3 - 3\tau_{3}]V_{brane}}{\dot{V}_{brane}}\nonumber \\&&
\bar{\tau} = \gamma\int_{0}^{t} d\tau' \frac{w}{\dot{w}} - \gamma
\frac{\sigma^{2}}{2}\label{qs2}
\end{eqnarray}
Using the above expression and doing the following calculations:
\begin{eqnarray}
\left\{\left[\left(\frac{\partial \bar{\tau}}{\partial \tau}\right)^{2} -
\left(\frac{\partial \bar{\tau}}{\partial
\sigma}\right)^{2}\right]\frac{\partial^{2}}{\partial \tau^{2}}+
\left[\left(\frac{\partial \rho}{\partial \sigma}\right)^{2} - \left(\frac{\partial
\rho}{\partial \tau}\right)^{2}\right]\frac{\partial^{2}}{\partial
\rho^{2}}\right\}TA=0\label{qs3}
\end{eqnarray}
we obtain:
\begin{eqnarray}
&& (-g)^{-1/2}\frac{\partial}{\partial
x_{\mu}}\left[(-g)^{1/2}g^{\mu\nu}\right]\frac{\partial}{\partial
x_{\upsilon}}TA=0\label{qs4}
\end{eqnarray}
where $x_{1}=\rho$, $x_{0}=\bar{\tau}$  and the metric elements become:
\begin{eqnarray}
&& g^{\bar{\tau}\bar{\tau}}\sim -\frac{1}{\beta^{2}}\left(\frac{w'}{w}\right)^{2}\frac{\left(1-\left(\frac{w}{w'}\right)^{2}\frac{1}{\sigma^{4}}\right)}
{\left(1+\left(\frac{w}{w'}\right)^{2}\frac{(1+\gamma^{-2})}{\sigma^{4}}\right)^{1/2}}=\nonumber
\\&&
-\frac{1}{\beta^{2}}\left(\frac{1+V_{brane}\dot{V}_{brane}^{-2}}{\frac{3\tau_{3} +[3 - 3\tau_{3}]V_{brane}}{\dot{V}_{brane}}}\right)^{2}\frac{\left(1-\left(\frac{\frac{3\tau_{3} +[3 - 3\tau_{3}]V_{brane}}{\dot{V}_{brane}}}{1+V_{brane}\dot{V}_{brane}^{-2} }\right)^{2}\frac{1}{\sigma^{4}}\right)}
{\left(1+\left(\frac{\frac{3\tau_{3} +[3 - 3\tau_{3}]V_{brane}}{\dot{V}_{brane}}}{1+V_{brane}\dot{V}_{brane}^{-2}}\right)^{2}\frac{(1+\gamma^{-2})}{\sigma^{4}}\right)^{1/2}}\nonumber
\\&&g^{\rho\rho}\sim -(g^{\bar{\tau}\bar{\tau}})^{-1}\label{qs5}
\end{eqnarray}
where we have assumed($\frac{\partial
TA}{\partial t} = \frac{\partial TA}{\partial \tau}= 2
\frac{\partial TA}{\partial \sigma}$).

Now, we can compare the elements with the line element of one
black $D3$-brane \cite{q21}:
\begin{eqnarray}
&& ds^{2}= D^{-1/2}\bar{H}^{-1/2}(-f
dt^{2}+dx_{1}^{2})+D^{1/2}\bar{H}^{-1/2}(dx_{2}^{2}+dx_{3}^{2})+D^{-1/2}\bar{H}^{1/2}(f^{-1}
dr^{2}+r^{2}d\Omega_{5})^{2},\nonumber
\\&&\label{qs6}
\end{eqnarray}
where
\begin{eqnarray}
&&f=1-\frac{r_{0}^{4}}{r^{4}},\nonumber
\\&&\bar{H}=1+\frac{r_{0}^{4}}{r^{4}}\sinh^{2}\alpha , \nonumber
\\&&D^{-1}=\cos^{2}\varepsilon + H^{-1}\sin^{2}\varepsilon, \nonumber
\\&&\cos\varepsilon =\frac{1}{\sqrt{1+\frac{\beta^{2}}{\sigma^{4}}}}.
\label{qs7}
\end{eqnarray}
Eqs. (\ref{qs5}) and (\ref{qs7}) lead to
\begin{eqnarray}
&&f=1-\frac{r_{0}^{4}}{r^{4}}\sim
1-\left(\frac{w}{w'}\right)^{2}\frac{1}{\sigma^{4}}=\nonumber
\\&& 1-\left(\frac{
\frac{3\tau_{3} +[3 - 3\tau_{3}]V_{brane}}{\dot{V}_{brane}}}{1+V_{brane}\dot{V}_{brane}^{-2}}\right)^{2}\frac{1}{\sigma^{4}},\nonumber
\\&&\bar{H}=1+\frac{r_{0}^{4}}{r^{4}}\sinh^{2}\alpha \sim 1+\left(\frac{w}{w'}\right)^{2}\frac{(1+\gamma^{-2})}{\sigma^{4}}= \nonumber
\\&& 1+\left(\frac{\frac{3\tau_{3} +[3 - 3\tau_{3}]V_{brane}}{\dot{V}_{brane}}}{1+V_{brane}\dot{V}_{brane}^{-2}}\right)^{2}\frac{(1+\gamma^{-2})}{\sigma^{4}}  \nonumber
\\&&D^{-1}=\cos^{2}\varepsilon + \bar{H}^{-1}sin^{2}\varepsilon\simeq1\nonumber
\\&&
\Rightarrow r\sim \sigma,r_{0}\sim
\left(\frac{w}{w'}\right)^{1/2},(1+\gamma^{-2})\sim \sinh^{2}\alpha
\label{qs8}
\end{eqnarray}
The temperature of the BIon system  is ${\displaystyle T=\frac{1}{\pi r_{0} \cosh\alpha}}$
\cite{w5}. As a result, the temperature of the brane-antibrane
system can be calculated as:
\begin{eqnarray}
&& T=\frac{1}{\pi r_{0}
\cosh\alpha}=\frac{\gamma}{\pi}\left(\frac{w'}{w}\right)^{1/2}\sim \nonumber
\\&& \frac{\gamma}{\pi}\left(\frac{1+V_{brane}\dot{V}_{brane}^{-2}}{\frac{3\tau_{3} +[3 - 3\tau_{3}]V_{brane}}{\dot{V}_{brane}}}\right)^{1/2}   \label{qs9}
\end{eqnarray}
The above equation shows that the temperature of thermal tachyonic black branes depends on the tachyonic potential and its changes with respect to time. By increasing the tacyonic potential, the temperature of the system grows and tends to large values. 

Like the previous section, we obtain the energy b using the energy-momentum tensor for the
black $D3$-brane \cite{w5} and write:
\begin{eqnarray}
&& V_{tot}= \int d\sigma T^{00} =
\int d\sigma \frac{\pi^{2}}{2}T_{D3}^{2}r_{0}^{4}(5+4\sinh^{2}\alpha)\sim \nonumber
\\&& 
\frac{\pi^{2}}{2}T_{D3}^{2}\left(\frac{w}{w'}\right)^{1/2}(9 +
\gamma^{-2})\sim \nonumber
\\&& \frac{\pi^{2}}{2}T_{D3}^{2}\left(\frac{\frac{3\tau_{3} +[3 - 3\tau_{3}]V_{brane}}{\dot{V}_{brane}}}{1+V_{brane}\dot{V}_{brane}^{-2}}\right)^{1/2}(9 +
\gamma^{-2})
\label{qs10}
\end{eqnarray}
Substituting the energy (\ref{qs10}) in equation (\ref{E}), we obtain:
\begin{eqnarray}
 && S_{BH}=4\pi \Big((V_{tot})^{2}-\frac{n}{2}\Big)\label{qs11}
\end{eqnarray}
The above equation shows that the spectrum of  the thermal tachyonic black branes depend on the tachyonic potential and its changes in terms of time. If the velocity of change in the tachyonic potential increases, branes move towards each other with high acceleration and  greater energy is produced in the system. This extra energy can be seen as the extra spectrum around black branes. 

\section{Results and discussion} 

In this research,  we have shown that by decreasing the separation between branes, the tachyonic energy increases. This causes  the quantum spectrum of the black hole to grow and the entropy increases. Also, we have found that the spectrum of  the thermal tachyonic black branes depend on the tachyonic potential and its change in terms of time. If the velocity of change in the tachyonic potential increases, branes move towards each other with high acceleration and  greater  energy is produced in the system. This extra energy can be seen as the extra spectrum around black branes.

\section{Conclusions} 

In this research, we have obtained the quantum spectrum of tachyonic black holes in brane-anti-brane systems. These black holes are built  from a pair of branes and anti-branes which are connected by a thermal tube. This tube is produced by a tachyonic potential between the branes. By decreasing the distance between the branes, the potential of the tachyons increase, and the energy of the system increases. Consequently, this black hole radiates extra energy and the quantum spectrum of black hole increases.\\

\noindent
{\bf Data Availability}\\

\noindent
No data was used to support this study.\\

\noindent
{\bf Conflicts of interest}\\

\noindent
The author declares that there is no conflict of interest regarding the publication of this paper.\\

\noindent
{\bf Funding Statement}\\

\noindent
This research received no specific funding, but was performed as part of the employment of the author with Mangosuthu University of Technology, South Africa. 


\end{document}